\documentclass[aps,prl,preprint,amssymb,floats,floatfix]{revtex4}
\usepackage{graphicx}
\usepackage{epic,eepic}
\usepackage{amsmath}
\begin{document}
\title{First-Principles Simulations of Inelastic Electron Tunneling Spectroscopy
of Molecular Junctions}
\author{Jun Jiang$^{1,2}$, Mathias Kula$^1$, Wei Lu$^2$ 
and Yi Luo$^1$\footnote{Corresponding author, Email: luo@theochem.kth.se}}
\affiliation{$^1$ Theoretical Chemistry, Royal Institute of Technology,
AlbaNova, S-106 91 Stockholm, Sweden}
\affiliation{$^2$ National Lab for Infrared Physics, Shanghai Institute of
Technical Physics, Chinese Academy of Sciences, China}
  
\input epsf
  
\vspace*{\baselineskip}
 
\begin{abstract}
A generalized Green's function theory is developed to simulate the 
inelastic electron tunneling spectroscopy (IETS) of molecular junctions. 
It has been applied to a realistic molecular junction with
an octanedithiolate embedded between two gold contacts 
in combination with the hybrid density functional theory calculations.  
The calculated spectra are in excellent agreement with 
recent experimental results. Strong temperature dependence of the
experimental IETS spectra is also reproduced. 
It is shown that the IETS is extremely sensitive to the intra-molecular
conformation and to the molecule-metal contact geometry.  
\end{abstract}
                                                                                          
\maketitle

Inelastic electron tunneling spectroscopy (IETS) has been recently applied 
to the molecular junctions\cite{kush04nano,wang04nano,yu04prl}. 
The measured spectra show well-resolved vibronic structures corresponding 
to certain vibrational normal modes of the molecule. It helps to understand 
the vibronic coupling between the charge carriers and nuclear motion of 
molecule. The IETS has also been found to be sensitive to the experimental 
setups, for instance, distinct difference has been revealed between 
the spectra of alkanethiol molecular junctions measured by 
Kushmerick et al. \cite{kush04nano} and Wang et al.\cite{wang04nano}. 
It is therefore expected that the IETS can be used as a tool to identify 
the geometrical structures of the molecule-metal interface, an important 
factor in the design and characterization of molecular devices.  
The basic theory of IETS has been known for many years\cite{wolf85}. 
Its extension to the molecular wires\cite{troi03jcp,asai04prl}, atomic
metal wires\cite{fred04prl} and molecular 
junctions\cite{troi03jcp,galp04nano,chen04nano}, however, has just appeared 
in the literature very recently, 
partly motived by the success of the experiments. The theoretical 
modellings are generally based on the Green's function theory at 
different levels, and have contributed to many fundamental understandings 
of the inelastic electron tunneling processes inside a molecular/atomic 
junction.  However, none of the theoretical studies has been able to 
make realistic comparisons with the experimental spectra.   

In this letter we present a new computational scheme based on our 
early developed quantum chemical approach for electron transport 
in molecular junction\cite{luo}.  The vibronic coupling is introduced 
by expanding the electronic wavefunction along different vibrational normal 
modes. We have applied this new method to the gold-octanedithiolate-gold 
junction. The electronic structures and vibrational modes are calculated using 
the hybrid density functional theory (DFT). The simulated spectra are in 
excellent agreement with the experimental ones. Our computational scheme 
can be easily applied to other systems. Theoretical simulations are
extremely useful for assigning the experimental spectra and to reveal 
many detailed informations that are not accessible in the experiments. 

\begin{figure}
\vspace*{-0.0cm}
\begin{center}
\epsfxsize=420pt\epsfbox{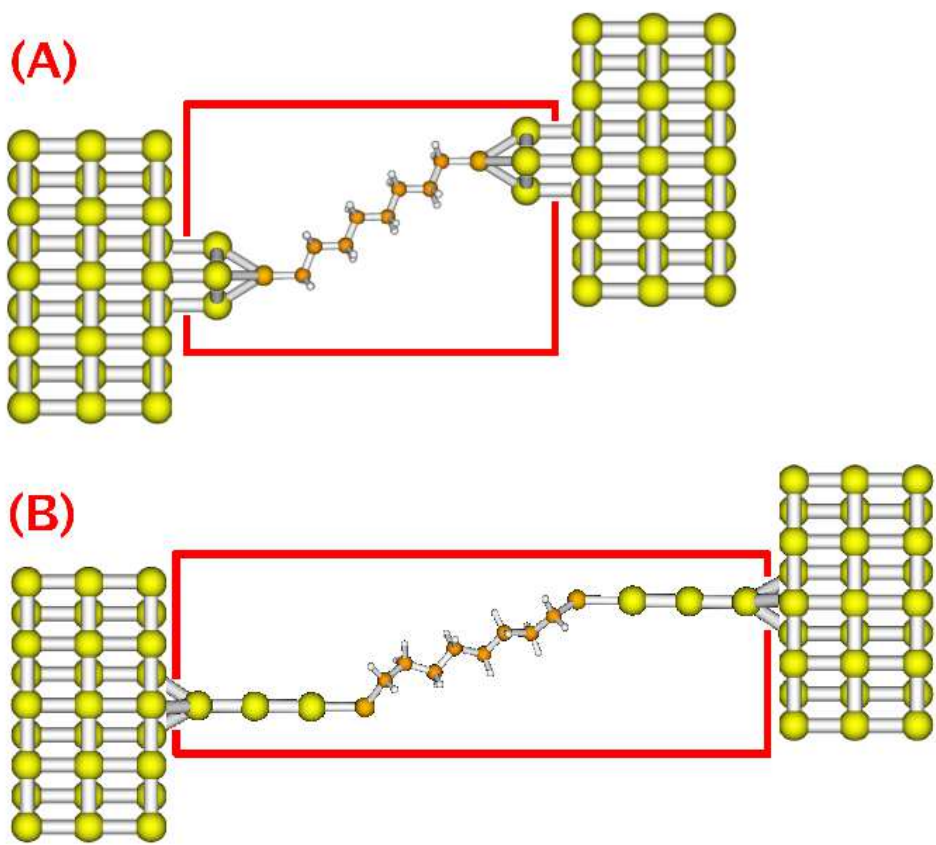} \vspace*{-0.0cm}
\hspace*{-0.0cm} \caption{Structures of the gold-octanedithiol-gold
junctions with triangle (A) and chain (B) local contacts.} \label{geo}
\end{center}
\vspace*{-0.0cm}
\end{figure}

Our approach is also based on the Green's function formalism. 
We divide  the molecular device into three parts, source, drain and 
extended molecule, as shown in Figure \ref{geo}.  The source and drain are 
described by an effective mass approximation (EMA), while the extended molecule 
is treated with the hybrid density functional theory. The extended molecule
is in equilibrium with the source and drain through the line up of their 
effective Fermi level. This approach is proved to lead to very good 
descriptions for the experimental results\cite{luo}.  The current 
density for the three-dimensional electrodes along the molcular
axis $z$ is given as\cite{luo} 
\begin{eqnarray}\label{cur3d1}
i_{SD}=
\frac{em^*k_B {\rm T}}{\hbar^3}
\int_{0}^{\infty}  \ln \frac{ 1+\exp ( \frac{E_f+eV_D-E_z}{k_B {\rm T}}) }{ 1+\exp( \frac{E_f-E}{k_B {\rm T}})} 
|T(V_D,Q)|^2 n^S n^D  {\rm d} E_z
\end{eqnarray}
where $n^S$ and $n^D$ are the density of states (DOS) of the
source and drain, respectively.
$E_f$ is the Fermi energy, and ${\rm T}$ is the
device working temperature.  $V_D$ is the external voltage. 
The transition matrix element from the source to the drain, 
$T(V_D,Q)T(V_D,Q)$, is dependent on the vibronal motion, $Q$, and 
can be written as\cite{luo}
\begin{eqnarray}\label{transab}
T(V_D,Q)= \sum_{J} \sum_{K} V_{JS}(Q) V_{DK}(Q) 
\sum_{\eta} \frac{ \langle
J(Q) \mid \eta (Q) \rangle \langle \eta(Q)  \mid K(Q)  \rangle }{z_{\eta}(Q)
-\varepsilon_{\eta}(Q)}
\end{eqnarray}
where $J$ and $K$ run over all atomic sites of the molecule, 
which are denoted as
1,2,...,$N$, site 1 and $N$ are two end sites of molecule that connect
with two electron reservoirs. $V_{JS}$ ($V_{DK}$) represents the coupling
between atomic site $J$ ($K$) and reservoirs S (D). Orbital $\mid \eta \rangle$
is the eigenstate of the Hamiltonian ($H_f$) of a finite system that
consists of the molecule sandwiched between two clusters of metal atoms:
$H_f\mid \eta \rangle=\varepsilon_{\eta} \mid \eta \rangle$.
The product of two overlap matrix elements $\langle
J \mid \eta \rangle \langle \eta \mid K \rangle$
represents the delocalization of orbital $\mid \eta \rangle $.
Here parameter $z_{\eta}$ is a complex variable, 
$z_{\eta}=E_{\eta}+i\Gamma^{JK}_{\eta}$,
where $E_{\eta}$ is the energy related to the external bias and Fermi level,
at which the scattering process is observed.
$\Gamma^{JK}_{\eta}$ is the escape rate which is determined by  
the Fermi Golden rule
\begin{eqnarray}\label{gamma}
\Gamma_{\eta}^{J K}(Q)
=\pi n^{S} V_{JS}^2 (Q) \mid \langle J(Q)  \mid \eta (Q) \rangle \mid^2
+ \pi n^{D} V_{DK}^2(Q)  \mid \langle \eta(Q)  \mid K (Q)  \rangle \mid^2
\end{eqnarray}
All the key parameters are obtained from the calculations of the 
finite metal-molecule cluster. The use of the finite system allows 
us easily including the vibronic coupling since the description of 
vibronic coupling in a molecular system has long been established. 
The nuclear motion dependent wavefunction can be expanded along 
the vibrational normal mode using a Taylor expansion. The IETS
experiment is often done at electronic off-resonant region. The adiabatic
harmonic approximation can thus be applied, therefore, only 
the first derivative like $\frac{\partial \Psi (Q) }{\partial Q_{a}}$ 
needs to be considered, where $Q_a$ is the vibrational normal mode 
$a$ of the extended molecule.  The working formula we have used for 
IETS calculations follows the same principle as the one given 
by Troisi, Ratner, and Nitzan\cite{troi03jcp}. 

We have calculated an octanedithiolate, $SC_{8}H_{16}S$, embedded between  
two gold electrodes through S-Au bonds.  The extended molecule consists of 
two gold trimers bonded with an octanedithiolate molecule. 
The gold trimers were tested in two basic conformations - a triangle and 
a linear chain, see Figure \ref{geo}.  For the triangle configuration, 
the sulfur atom is placed above the the middle of the triangle, resembling 
the hollow site of a Au(111) surface. Geometry optimization and electronic 
structure calculations have been carried out for the extended molecules 
of different configurations at hybrid DFT B3LYP level \cite{B3LYP} 
using Gaussian03 program package with the LanL2DZ basis set \cite{gau03}. 
In the case of the triangle configuration, two different schemes have
been used for the geometry optimization. The first geometry is obtained by 
optimizing octanedithiol in gas phase and then replace the terminal 
hydrogens with the gold contacts without further optimization (Tr1). 
The second geometry is obtained by optimizing the first geometry with 
the same fixed gold and sulfur distance (Tr2), which is also the approach 
for the chain configuration (Ch1). The S-Au distance in all calculations 
is fixed to 2.853 {\AA}. The QCME program\cite{qcme} has been employed for 
all the IETS calculations.       

\begin{figure}[ht]
\begin{center}
\epsfxsize=320pt\epsfbox{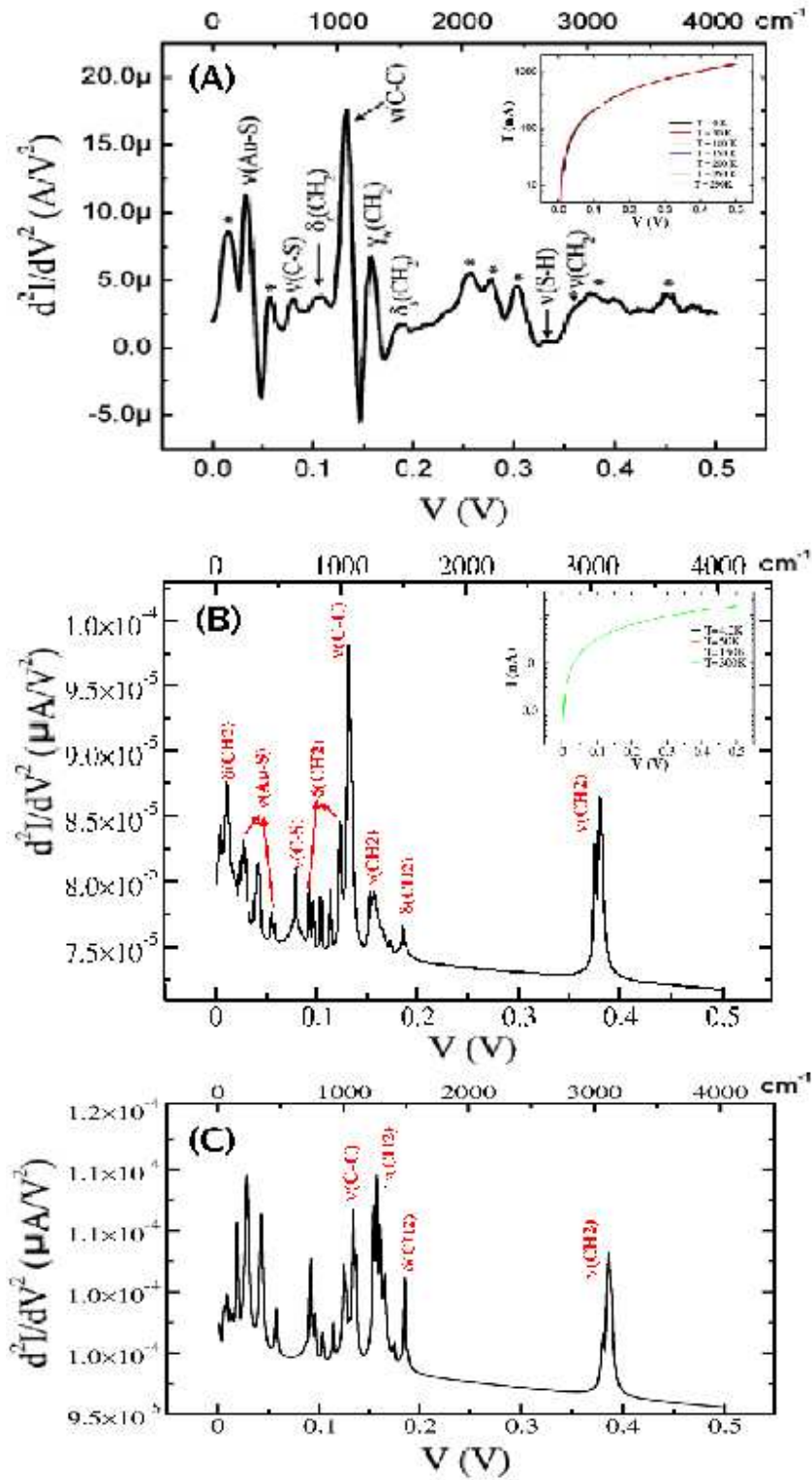} \vspace*{-0.0cm}
\hspace*{-0.0cm} \caption{Inelastic electron tunneling spectrum of
the octanedithiol junction from (A) experiment\cite{wang04nano},
(B) calculation for the Tr1 configuration and (C) calculation for the Tr2
configuration. The I-V curves are given in the inserts of
(A) and (B).  The working temperature is 4.2 K.} \label{expcalvib}
\end{center}
\end{figure}

The calculated IETS of the octanedithiolate junction with the triangle 
gold trimers are shown in Figure \ref{expcalvib}, together with the 
experimental spectrum of Wang et al. at temperature 4.2 K \cite{wang04nano}. 
One can clearly see that the calculated result for Tr1 configuration is 
in excellent agreement with the experiment. The calculations do not only 
reproduce all the major spectral features observed in the experiment, 
but also provide very detailed features that are smeared out by the 
background due to the encasing Si$_3$N$_4$ marked with stars \cite{wang04nano}.
Both theory and experiment shows that the intensity of the vibronic feature 
follows the order: 
$\nu$(C-C) (131mV) $>$ $\gamma$(CH$_2$) (155mV) $>$ $\delta$(CH$_2$) (185mV).
It should also be mentioned that theory and experiment agrees well 
for the current-voltage characteristics (I-V) which are determined mainly 
by the elastic scattering, see the inserts of the Figure \ref{expcalvib} 
(A) and (B).    

The IETS calculated with the Tr2 configuration, Figure \ref{expcalvib} (C), 
shows also very rich structures. However, its spectral profiles are quite 
different from those obtained from the Tr1 configuration, as well as 
the experiment of Wang et al\cite{wang04nano}.  For instance, the spectral 
peak at 155 mV ($\gamma$(CH$_2$) mode) has the largest intensity, 
instead of the mode of $\nu$(C-C) at 131 mV.  The Tr1 and the Tr2 
configurations have the same Au-S bondings, but slightly different
intra-molecular conformations. The sensitivity of the IETS with respect 
to the molecular geometry is really high. It can thus be 
concluded that the molecular geometry and contact configuration of the 
device in the experiment of Wang et al\cite{wang04nano} is very close 
to our Tr1 configuration. 

\begin{figure}
\vspace*{-0.0cm}
\begin{center}
\epsfxsize=420pt\epsfbox{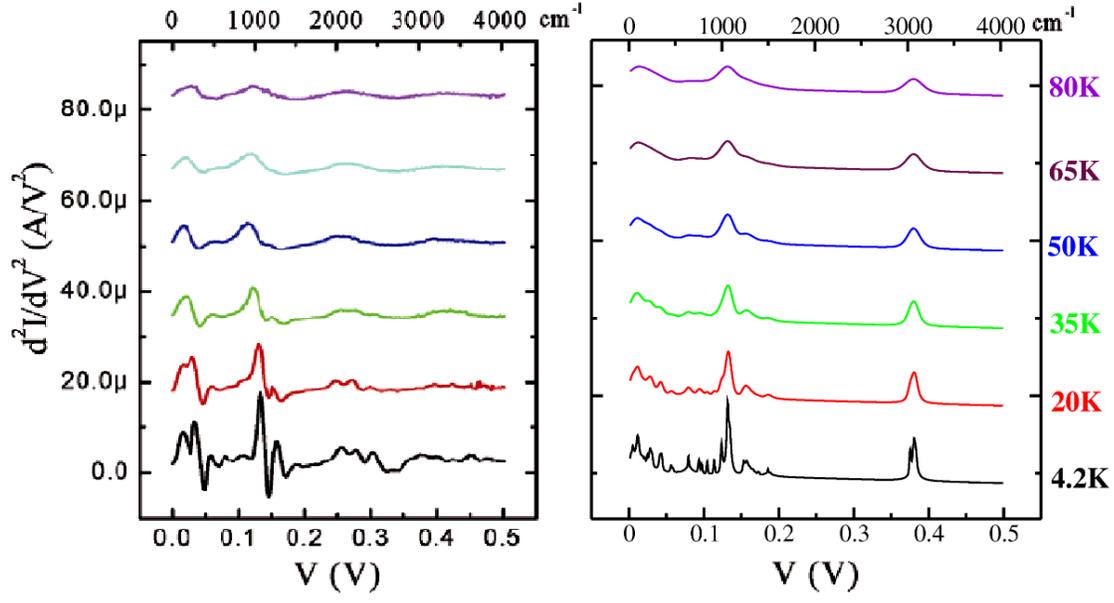} \vspace*{-0.0cm}
\hspace*{-0.0cm} \caption{Temperature dependent IETS of the octanedithiolate
junction from (A) experiment\cite{wang04nano}, and (B) calculation
for the Tr1 configuration. The intensity is in arbitrary unit.}
\label{allvibtemp}
\end{center}
\vspace*{-0.0cm}
\end{figure}

The calculated temperature dependence of the IETS of the Tr1 configuration 
is shown in Fig.\ref{allvibtemp} together with the experimental results of 
Wang et al\cite{wang04nano}. The agreement between the theory and the 
experiment is more than satisfactory. The evolution of the spectral bands 
upon the increase of the temperature is the same for both the experiment
and the calculation. As an example, in both cases, the peak for 
mode $\delta$(CH$_2$) at 185 mV disappears at 35K, and the peak for 
mode $\gamma$(CH$_2$) at 155 mV becomes invisible at 50K. Our simulations 
indicate that the observed temperature dependence is mainly due to the 
changes of the Fermi distribution. 

In Table \ref{vibdata}, the assignments of the vibronic bands observed
in theoretical and experimental spectra are shown. It can be seen that
the vibrational frequencies given by the B3LYP calculations are in very good
agreement with the experiments. Our computational scheme also allows to
calculate the spectral linewidth directly, which is determined by the orbital
characters and molecule-metal bonding, see Eq. \ref{gamma}.
The calculated full width at half maximum (FWHM) for the spectral profile of
mode $\nu$(C-C) at 132 mV is found to be around 4.3 meV, in good agreement 
with the experimental result of 3.73$\pm$0.98 meV\cite{wang04nano}. However, 
it is also noticed that such a band is attributed from several modes of the 
same character with vibration frequencies 1012, 1038, 1061, 1069, 1084, 1090, 
1097, and 1100 cm$^{-1}$, respectively, covering a range of 11.0 meV. The 
overlapping between different vibration modes makes it impossible to 
determine the actual intrinsic lindwidth of the spectral profile from a 
single vibration mode.  

\begin{figure}
\vspace*{-0.0cm}
\begin{center}
\epsfxsize=420pt\epsfbox{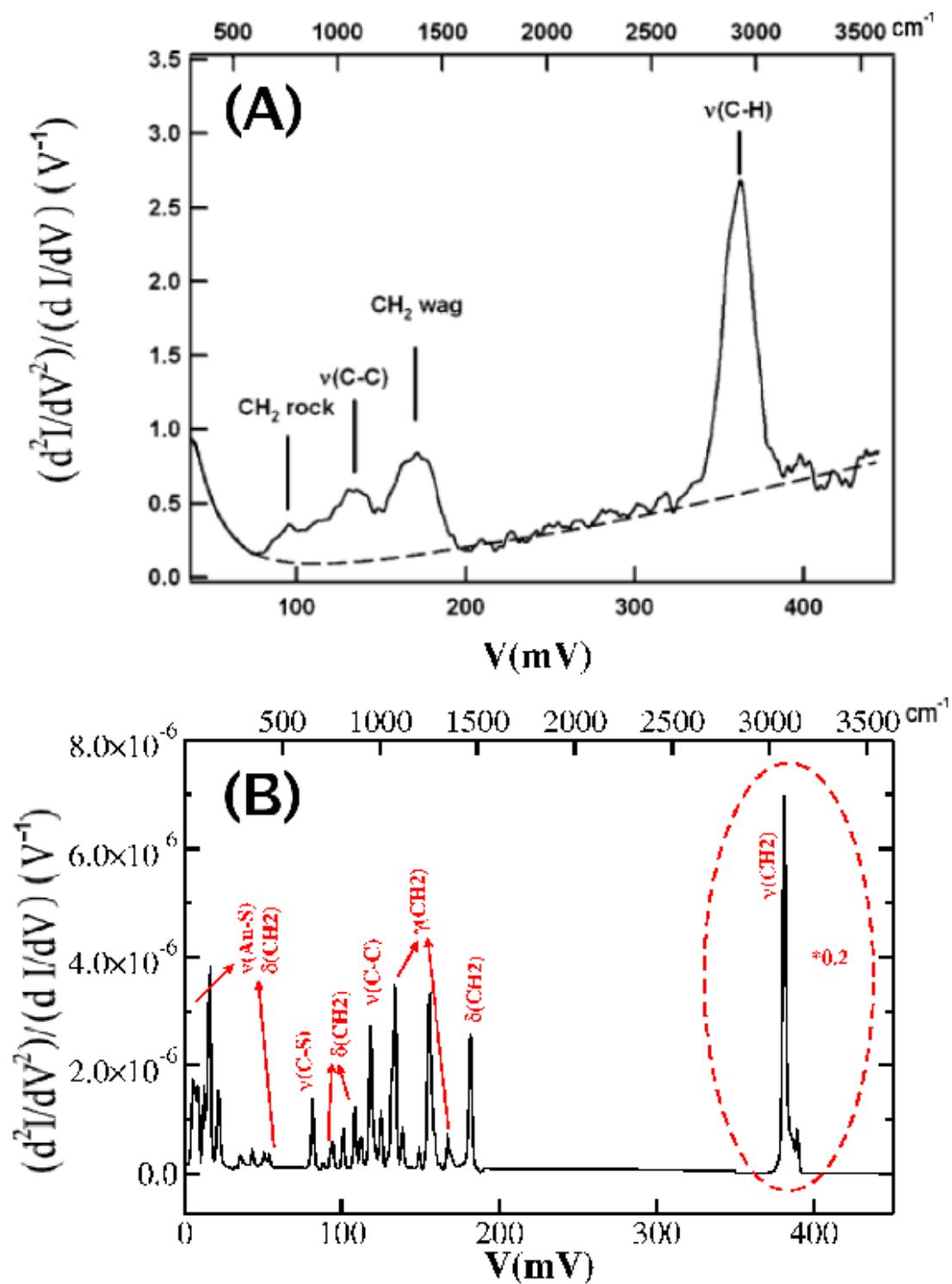} \vspace*{-0.0cm} \hspace*{-0.0cm} \caption{Inelastic electron tunneling spectroscopy from
(A) experiment for C11 monothiol \cite{kush04nano} and (B) calculation for
the Ch1 configuration.  The working temperature is  4.2 K.}
\label{chainexpcalvib}
\end{center}
\vspace*{-0.0cm}
\end{figure}

We have calculated the IETS of the gold chain configuration (Ch1), shown 
in Figure \ref{chainexpcalvib}(B), to examine the dependence of the IETS 
on the molecule-metal bonding structure. Indeed, the IETS of Ch1 shows a 
distinct difference in spectral intensity distribution from those of the 
Tr1 and the Tr2 configurations.  The spectral peak of mode $\delta$(CH$_2$) 
at 382 mV has become the absolute dominate feature in the spectrum. 
Furthermore, the intensities of different spectral features follow the 
order of $\nu$(C-C) (132 mV) $<$ $\gamma$(CH$_2$) (172 mV) $
<$ $\delta$(CH$_2$) (382 mV), completely different from the 
results of the Tr1 and the Tr2 configurations. The Au-S bonding
structure of the Ch1 configuration differs from that of the 
Tr1 and the Tr2, resulting a noticeable difference in 
their spectral profiles for the mode $\nu$(Au-S) around 
40 mV. The changes in molecular conformations seems 
to be the major cause for the large difference in the spectral intensity 
distributions of two devices. It is found that the molecule in 
the Ch1 configuration is twisted around the otherwise linearly 
oriented molecular back-bone in the Tr1 configuration. 
It is interesting to note that the spectrum of the
Ch1 configuration resembles the experimental IETS of an alkanemonothiol 
molecule, HS(CH$_2$)$_8$H (C11)\cite{kush04nano}, quite well, as clearly 
demonstrated in Figure \ref{chainexpcalvib}. The molecule-metal bonding 
structure of the C11 is very different from that of octanedithiol. 
Such a difference should be reflected by the spectral profiles of the 
Au-S modes at the low energy region. The large difference in the experimental 
spectral intensity distribution \cite{kush04nano,wang04nano} related to 
the molecular vibration modes implies that the molecular conformations 
in two experimental setups is very different. The molecular back-bone of 
the octanedithiolate junction in the device of Wang et al.\cite{wang04nano} 
should be linear, while it is slightly twisted for the C11 in the 
device of Kushmerick et al\cite{kush04nano}.   

In conclusion, we have proposed a new computational scheme that 
is capable of describing the IETS of molecular junctions with unprecedented 
accuracy. Our first-principles calculations provide reliable 
assignments for the experimental spectra and reveal important details 
that are not accessible in the experiment.  We have also demonstrated 
that the IETS is a powerful characterization tool for molecular devices.  

Acknowledgments: This work was supported by the Swedish Research Council (VR),
the Carl Trygger Foundation (CTS),
and Chinese state key program for basic research(2004CB619004).

\newpage
\begin{table}
\begin{center}
\caption{Assignments of the vibrational modes observed in the IETS of the
octanedithiolate junction. The calculated peak positions (in meV),
the full width at half maxmimum (FWHM) of a single spectral line (in meV),
the beginning and the ending of a spectral band (in cm$^{-1}$), and the
total width of the band $\Delta$ (in meV) are given. The experimental
results of Wang et al. \cite{wang04nano} are also listed for comparison.}
\vspace*{0.1cm}
\begin{tabular}{ccccccc}
\hline
\hline
Mode&\multicolumn{3}{|c|}{Peak(meV)} & \multicolumn{3}{|c}{Band} \\
\cline{1-7}
    &\multicolumn{1}{c|}{Exp.$^a$} & \multicolumn{1}{c|}{Cal.}& \multicolumn{1}{c|}{FWHM} & \multicolumn{1}{c|}{begin} & \multicolumn{1}{c|}{end} & \multicolumn{1}{c}{$\Delta$} \\
\cline{1-7}
\hline
$\delta$(CH$_2$)&--&12&5.3&93&138&5.4\\
\hline
$\nu$(Au-S)&34&28&4.8&228&250&2.7\\
&--&44&4.2&335&353&2.3\\
&--&57&3.3&453&474&2.6\\
\hline
$\nu$(C-S)&79&79&2.2&639&652&1.6\\
\hline
$\delta$(CH$_2$)&--&95&5.1&742&783&1.7\\
&104&104&1.1&841&841&0.0\\
&--&114&1.1&917&917&0.0\\
&--&124&--&991&998&1.0\\
\hline
$\nu$(C-C)&132&132&4.3&1012&1100&11.0\\
\hline
$\gamma$(CH$_2$)&159&155&13.9&1225&1402&22.0\\
\hline
$\delta$(CH$_2$)&180&185&7.2&1484&1536&6.4\\ \hline
$\nu$(CH$_2$)&354&379&8.4&3012&3146&17.0\\
\hline
\hline
\end{tabular}
\label{vibdata}
\end{center}
$^a$ Ref. \cite{wang04nano}\\
\end{table}
                                                                                
\newpage

\end{document}